# Deciphering the AI Economy: A Mathematical Model Perspective

Davit Gondauri[1], Mikheil Batiashvili[1] & Nino Enukidze[1]

[1] Business and Technology University, Tbilisi, Georgia

Correspondence: Davit Gondauri, Business and Technology University, Tbilisi, Georgia. E-mail: dgondauri@gmail.com

## Abstract

The economy in the modern world is greatly influenced by artificial intelligence (AI). The purpose of this paper is to determine the impact of AI quantitative relationships on the country's economic parameters, including GDP per Capita. Historical data analysis is used in the research. A new mathematical algorithm for the magnitude of a vector of technological level and AI factors has been developed. The study calculated the economic effect of AI on GDP per Capita. As a result of the analysis, it was revealed that there is a positive Pearson correlation between growth. On AI and GDP per Capita, that is, to increase GDP per Capita by 1%, an average increase of 23.9% in AI is required.

## 1. Introduction

The development of innovation, technology and science has radically changed the way people live. A lot of things have been rethought, a completely new opportunity has been created, which people have already started to use. The result of such rapid development is the unique human creation - artificial intelligence. This human achievement is slowly qualitatively changing human approaches, to scientific results and to the function and purpose of man himself. Artificial intelligence, like a person, "spawns" in the country and creates capital. It is the latter raises the question, now, in the 21st century, what is intellectual capital and what elements does it consist of today and In general, what does the economy of the 21st century mean, what correlation does artificial intelligence have with the country's economy and GDP per capita?

In the study, we aimed to create an appropriate mathematical algorithm that would enable us to determine the technology level of the countries presented in the study, where the indicators are the country's innovation index, research and development costs, information technology exports, high technology exports, and patent applications by residents. After that, the authors continue the research by developing a vector model whose components are such indicators of artificial intelligence as Technological Development, AI Adoption rate, AI Workforce, AI Productivity, AI Market Demand, and AI Regulatory Environment. After deriving the vector algorithm, its regression analysis on GDP per capita was calculated. Reliability of regression analysis is tested by p-value.

Artificial intelligence will affect a certain percentage of existing jobs in developed economies, although it should be noted that this figure will be much smaller in developing and low-income countries. Artificial intelligence may negatively affect half of the jobs mentioned, but will have a positive effect on the remaining jobs due to increased productivity. For example, maybe a certain category of people will lose their jobs completely, which is certainly not good, or artificial intelligence will help them do their jobs more efficiently, which will increase people's income. Although artificial intelligence will initially have less negative impact on low-income countries, these countries will benefit less from new technologies. This situation can further widen the digital divide and income distribution between countries.

## 2. Literature Review

Looking at the trends in the use of automation, artificial intelligence, and robotics throughout history, it has been suggested that this may affect the labor market and lead to job losses. However, automation helps increase productivity through efficient use of electrical technology and information technology. Slow economic growth in recent years underscores the importance of AI's impact on the economy (Furman & Seamans, 2019).

In 2019 study by the National Bureau of Economic Research found that artificial intelligence is having a significant impact on the economy. Based on various statistical data, there is an assumption that the growth of artificial intelligence activity may have a mixed impact on the labor market, especially in the short term (Furman & Seamans, 2019).

The circumstances mentioned above, with some studies explaining that AI poses a threat of job loss for workers who cannot adapt to new technologies. In order to avoid these threats, it is necessary to develop an educational policy adapted to the labor market and invest in education (Gondauri & Batiashvili, 2023).

The rapid and far-reaching nature of the Fourth Industrial Revolution is driven by new technological advances, such as artificial intelligence, and affects the entire economic system. Since the 1950s, the use of AI to augment human intelligence in the face of cognitive limitations has been introduced. However, it did not gain attention from the business side until the Internet of Things emerged. Modern AI systems, along with their development, have become the driving force in the digital economy (Hang & Chen, 2022).

There has been a significant trend since 1950, reflected in the long-term decline in male labor force participation from 98% to 89% between 1950 and 2016, indicating the complexity of new skills and changing occupations. In this situation, it is crucial to acquire new skills to enable continuity of employment. The challenge of artificial intelligence is the rapid changes that may lead to disruption of the potential workforce, putting the need for retraining policies on the agenda (Furman & Seamans, 2019).

AI, as a driver of the digital economy, increases revenue by improving employee productivity, surveying customers and offering diverse resources, but its lack of interpersonal nature and management prevents it from realizing its full potential. For example, AI works with customers at the initial stage of purchases, responds to simple customer requests, detects fraud, receives loans, orders and develops innovative products. A large number of business representatives have already invested in the use of artificial intelligence. However, only 40% claim to have benefited from the use of AI. For this, it is necessary to determine how artificial intelligence creates advantages in the digital economy and what barriers prevent it from realizing its full potential (Hang & Chen, 2022).

It is known that the conceptual approaches used by economists to determine the impact of artificial intelligence on the labor market were the task approach in labor economics and the Acemoglu-Restrepo growth (AR) model. The AR model includes a task approach and its central equation is the production function for task automation and provided only by labor (Gries & Naudé, 2022).

The digital economy plays a crucial role in the development of the labor market, and the measurement and calculation of its mathematical algorithm is important for the modern world (Gondauri et al., 2023).

In 2024 Jie Sun study shows that the rapid development of big data technology has led to the creation of an artificial intelligence economy, which may benefit China's economic transformation and international competitiveness (Sun, 2024). The mathematical model of the artificial intelligence economy presented in the article is based on the theoretical foundations of financial mathematics and addresses the unstable development of China's artificial intelligence economy by building a theoretical framework for investment risk management. The convergence efficiency of the model is higher than that of the capital asset pricing model and the financial derivatives pricing model. The model used in the study will solve financial optimization problems, including portfolio, stock forecasting, risk management and other areas, providing optimization of decisions (Sun, 2024).

Financial mathematics is used to develop financial economics, which focuses on solving problems in the financial industry using mathematical applications. Financial problem solving involves data analysis and mathematical knowledge that increases efficiency and performance levels (Sun, 2024. For economic analysis, it is important to develop financial mathematical models that allow quantitative analysis.

The technological support of AI becomes part of social and economic activities, contributing to the development of a sustainable economy, even as it changes production structures and consumption patterns (Qin et al., 2023).

Artificial intelligence affects the economy in various aspects - automation of production, the function of generating ideas, the singularity and the intersection of artificial intelligence with firms and economic growth (Gondauri et al., 2024).

Last year, the researchers developed a geometric mean algorithm of various indices, resulting in the creation of a single index of the digital economy, which itself correlates with GDP (Gondauri et al., 2023).

## 3. Research Methodology

The first stage of the methodology in the research refers to the creation of a mathematical formula that measures the technological level of the countries represented in the research, the indicators of which are the country's innovation index, research and development costs, information technology exports, high technology exports. , and patent applications by residents. The second phase of the research concerns the development of a vector model with its own indicators such as technological development, AI adoption rate, AI workforce, AI productivity, AI market demand and AI regulatory environment. After deriving the vector algorithm, its regression analysis on GDP per capita was calculated. Reliability of regression analysis is tested by p-value.

To calculate the technological level, the authors used the weighted average data of the above-mentioned indicators for the years 2011-2022 for the following countries - Georgia, Israel, Armenia, Azerbaijan, Turkey, USA, France and Germany, To calculate the technological level of the country, the authors developed a mathematical formula of the geometric mean:

$$\sqrt[n]{Factor1 * factor2 * factor3 * \ldots * Factor(n)} \quad (1)$$

where the factors presented in the first step of the methodology are entered as factors.

In the second stage of the research, the authors calculated the magnitude of the vector with the appropriate algorithm:

$$\sqrt{factor1^2 + factor2^2 \ldots + factor(n)^2} \quad (2)$$

The following factors were used to calculate the magnitude of the vector in the formula:

• Technological Development - We make a distinction between the technological level and the decay term.

• AI Adoption rate - Differences in AI investment change and regulatory effects over time.

• AI Workforce Dynamics - We measure AI based on changes in investment, AI adoption and natural attrition.

• AI Productivity - This is a function of both technology level and speed of AI adoption.

• AI Market Demand - Market demand changes over time based on differences in AI productivity and decay time.

• AI Regulatory Environment - The difference between the terms AI adoption and decay affects the regulatory environment.

In the third stage of the research, the authors investigated the regression analysis of the gross domestic product of Georgia per person and the magnitude of the vector (2011-2022 years). To obtain the regression, the logarithms of the two upper exponents included in the third stage of the research were calculated using the Nepper base, after which the regression coefficient was obtained:

$$b_1 = \frac{\sum_{i=1}^{n}(x_i-\bar{x})(y_i-\bar{y})}{\sum_{i=1}^{n}(x_i-\bar{x})^2} \quad (3)$$

After the regression coefficient, the authors calculated the Pearson correlation coefficient:

$$r = \frac{\sum(x-\bar{x})(y-\bar{y})}{\sqrt{\sum(x-\bar{x})^2 \sum(y-\bar{y})^2}} \quad (4)$$

and the coefficient of determination – $(5)=(4)^2$

And finally, to test the correctness of the hypothesis

$$P = 2 \times P(T > |t|) \quad (5)$$

Where T is the t-distribution, and t is the t-statistic.

## 4. Research Results and Discussion

In the results of the research, their technological level was determined for several countries, to calculate which we used the data presented in Table 1, which represents the average indicator for the years 2011-2022 according to the countries. To calculate the technological level, we took statistics from website www.theglobaleconomy.com.

Table 1. Technological development factors

| Technological Development Factors | Georgia | Israel | Armenia | Azerbaijan | Turkey | USA | France | Germany |
|---|---|---|---|---|---|---|---|---|
| Technology level | 6.0 | 95.6 | 7.4 | 4.1 | 33.4 | 326.8 | 166.0 | 222.2 |
| Innovation index | 33.5 | 54.4 | 33.9 | 28.8 | 37.0 | 60.2 | 53.4 | 57.0 |
| Research and development expenditure, percent of GDP | 0.3 | 4.1 | 0.2 | 0.3 | 0.8 | 2.7 | 2.2 | 2.7 |
| Information technology exports, percent of total goods exports | 0.5 | 11.1 | 0.6 | 0.0 | 2.4 | 11.7 | 5.2 | 5.9 |
| High technology exports (million U.S. dollars) | 21 | 11794 | 46 | 25 | 3504 | 173106 | 105533 | 197273 |
| High tech exports, percent of manufactured exports | 2.8 | 21.0 | 5.7 | 4.5 | 2.8 | 21.5 | 24.9 | 16.4 |
| Patent applications by residents | 209 | 1263 | 133 | 200 | 2087 | 169515 | 13255 | 41517 |

Table 1 shows the level of technology in different countries. According to the table, the USA is in the first place regarding technological development.

We continue the research results by deriving the magnitude of the vector, which is presented in Table 2.

Table 2. Vector factors

| Vector Factors | Georgia | Israel | Armenia | Azerbaijan | Turkey | USA | France | Germany |
|---|---|---|---|---|---|---|---|---|
| Technological Development | 6.01 | 95.65 | 7.37 | 4.11 | 33.40 | 326.85 | 165.98 | 222.15 |
| AI Adoption | 48.0% | 55.0% | 55.0% | 55.0% | 55.0% | 50.0% | 48.0% | 48.0% |
| AI Workforce Dynamics | 0.52 | 1.22 | 0.39 | 1.44 | 9.60 | 47.38 | 8.87 | 12.42 |
| AI Productivity | 62.6% | 62.6% | 62.6% | 62.6% | 62.6% | 62.6% | 62.6% | 62.6% |
| Market Demand | 61.3% | 61.3% | 61.3% | 61.3% | 61.3% | 61.3% | 61.3% | 61.3% |
| Regulatory Environment | 48.7% | 48.7% | 48.7% | 48.7% | 48.7% | 48.7% | 48.7% | 48.7% |
| Magnitude of a vector | 6.13 | 95.66 | 7.47 | 4.50 | 34.77 | 330.27 | 166.22 | 222.50 |

According to the countries, with the vector volume obtained in Table 2, Georgia shows a relatively low level of technological level and adoption of artificial intelligence.

After that, on the example of Georgia, we examined the regression analysis of the vector size (Table 2) and GDP per capita on the example of Georgia and got the results:

1. Regressive (slope) coefficient 23.9% - This means that to increase GDP per capita by 1%, the vector model must increase by 23.9%;

2. As a result of the research, the determination rate was 77.3%

3. And the P value is (0.0435).

Regression coefficient:

•If the regression coefficient is positive (eg, +23.9%), it indicates a positive correlation between the two variables.

If one variable increases, the other variable will increase by the specified percentage.

• If the regression coefficient is negative (eg -23.9%), this indicates a negative correlation. If one variable

increases, another variable in the algorithm will decrease by the specified percentage.

Coefficient of determination:

A coefficient of determination with a value of 77.3% indicates that approximately 73% of the sensitivity of the dependent variable in the study is explained by the independent variable. The obtained value implies that the presented model fits the data well.

P-value:

The low p-value of 0.0435 obtained in the study is below conventional significance levels (ie, 0.05). The result clearly indicates that the relationship between the variables is statistically significant.

Clarifying and analyzing the practical implications of the research is critical for policymakers and businesses to effectively use research to inform their strategies.

Policy development: For example, insights into the impact of regulatory changes on AI adoption and performance can help policymakers design regulatory frameworks that foster innovation, ensure ethical AI development and deployment, and foster economic growth.

Business Strategy: By understanding how regulatory changes may impact AI adoption and efficiency, businesses can proactively adjust their strategies to navigate regulatory challenges and capitalize on opportunities.

Risk management: includes assessing the potential impact of regulatory changes on business operations, intellectual property rights, data privacy, and cyber security.

Stakeholder Engagement: Policymakers and businesses can use research findings to engage with stakeholders, including industry associations, advocacy groups, and the public.

Capacity building: This may include providing training and education programs to policymakers, regulators, industry professionals, and other relevant stakeholders to enhance understanding of AI issues and best practices.

## 5. Conclusion

Artificial intelligence is a completely new opportunity for the economy. This latest capital allows It allows us to dramatically increase economic resources. Various data show that the countries whose economies are taking part in the creation of the artificial Having intelligence, i.e. artificial intelligence – capital Companies are the fastest growing and undergoing transformation. That is why it is necessary to study The role of artificial intelligence-capital. Let us introduce different methodologies to evaluate its role in the economy and apply it to the evolution of humanity.

In conclusion, it should be noted that the study places crucial importance on the growth of the technological level in the country and the importance of introducing artificial intelligence to accelerate economic growth in the country. From the results of the study (Table 1), it can be seen that the USA is the leader in technological development among the presented countries, and the magnitude of the vectors of the technological level of the countries investigated in Table 2 indicates a relatively low level of technological progress and adoption of artificial intelligence in Georgia compared to other countries, such as Israel, the USA, and Germany.

Overall, the study provides a basis for policymakers, businesses, and researchers to formulate informed strategies and decisions aimed at promoting innovation and sustainable economic progress.

**Informed consent**

Obtained.

**Ethics approval**

The Publication Ethics Committee of the Canadian Center of Science and Education.

The journal and publisher adhere to the Core Practices established by the Committee on Publication Ethics (COPE).

**Provenance and peer review**

Not commissioned; externally double-blind peer reviewed.

**Data availability statement**

The data that support the findings of this study are available on request from the corresponding author. The data are not publicly available due to privacy or ethical restrictions.

**Data sharing statement**

No additional data are available.

**Open access**

**Copyrights**